\documentclass[prl,twocolumn,showpacs,preprintnumbers,amsmath,amssymb]{revtex4}
\usepackage{graphicx}
\usepackage{dcolumn}
\usepackage{bm}

\begin{document}

\title{Above-room-temperature ferromagnetism in  half-metallic Heusler
     compounds  NiCrP, NiCrSe, NiCrTe  and NiVAs: A first--principles study}

\author{E.~\c Sa\c s\i o\~glu, L. M.  Sandratskii and  P. Bruno}

\affiliation{Max-Planck Institut f\"ur Mikrostrukturphysik,
D-06120 Halle, Germany }

\date{\today}

\begin{abstract}

We study the interatomic exchange interactions and Curie
temperatures in  half--metallic semi Heusler compounds NiCrZ
(Z$=$P, Se, Te) and NiVAs. The study is performed within the
framework of density functional theory. The calculation of
exchange parameters is based on the frozen--magnon approach. It is
shown that the  exchange interactions in NiCrZ vary strongly
depending on the Z constituent. The Curie temperature, T$_c$, is
calculated within the mean field and random phase approximations.
The difference between two estimations is related to the
properties of the exchange interactions. The predicted Curie
temperatures of all four systems are considerably higher than room
temperature. The relation between the half-metallicity and the
value of the Curie temperature is discussed. The combination of a
high spin-polarization of charge carriers and a high Curie
temperature makes these Heusler alloys interesting candidates for
spintronics applications.

\end{abstract}


\maketitle

\section{introduction}
\label{intro}

An important current problem on the way to the practical use of
the spin-transport in semiconductor devices is the fabrication of
the materials that make possible the injection of spin-polarized
electrons into a semiconductor at room temperature \cite{ohno}.
One of the classes of systems promising to supply materials with
necessary combination of properties is the half--metallic
ferromagnets, i.e., ferromagnetic systems where one spin channel
is metallic and other is semiconducting. The half-metallicity
leads to the 100\% spin-polarization of the electron states at the
Fermi level $E_F$ \cite{Zutic,deBoeck}. The half--metallic
ferromagnetism (HMF) was discovered by de Groot \textit{et al.}
when studying the band structure of semi Heusler compound NiMnSb
\cite{deGroot}. Ishida \textit{et al.}  have proposed that also
the full-Heusler alloys of the type Co$_2$MnZ, (Z=Si,Ge), are
half-metals \cite{Ishida}. Since then a number of further systems
were predicted to be half-metallic. Among them binary magnetic
oxides (CrO$_2$ and Fe$_3$O$_4$), colossal magnetoresistance
materials (Sr$_2$FeMoO$_6$ and La$_{0.7}$Sr$_{0.3}$MnO$_3$)
\cite{Soulen}, diluted magnetic semiconductors
(Ga$_{1-x}$Mn$_x$As)  and zinc-blende compounds MnAs and CrAs
\cite{Freeman,Akai,Akinaga}.

The Heusler alloys form a particularly interesting class of
materials since they are characterized by a high Curie temperature
and good crystallographic compatibility with the lattices of
semiconductors used industrially
\cite{Webster,lattice_match_1,lattice_match_2}. The main body of
previous first-principles studies was focused on the investigation
of the properties of the semiconducting gap in the minority-spin
channel. Galanakis \textit{et al.} considering NiMnSb have shown
that the gap arises from the interaction between the $d$-orbitals
of the Ni and Mn atoms leading to the formation of bonding and
antibonding states separated by a hybridizational gap
\cite{GalanakisHalf}. Chioncel \textit{et al.} have demonstrated
by the example of NiMnSb that the electron-magnon interaction can
lead to the appearance of nonquasiparticle states in the
half-metallic gap \cite{Chioncel}.

For a number of Heusler alloys it was shown that half--metallicity
is preserved under tetragonalization of the crystal lattice
\cite{Block} and application of the hydrostatic pressure
\cite{Picozzi-Gala}. Mavropoulos \textit{et al.} studied the
influence of the spin-orbit coupling on the spin-polarization at
the Fermi level and found the effect to be very small
\cite{Mavropoulos} that corresponds to a small orbital moment
calculated by Galanakis \cite{GalanakisOrbit}. Orgassa and
collaborators have shown that the half-metallic gap decreases with
increasing disorder \cite{Orgassa}. Picozzi and collaborators and
Miura \textit{et al.} demonstrated that different types of
structural defects have different influence on the
half-metallicity \cite{Orgassa,Picozzi2004,disorder}.

An important part of theoretical efforts is a first-principles
design of new half--metallic Heusler alloys. Galanakis studied the
appearance of half-metallic ferromagnetism in quaternary Heusler
alloys \cite{quaternary}. Xing \textit{et al.}  have predicted the
half--metallic ferromagnetism in NiCrZ (Z$=$P, Se, Te) and NiVAs
semi Heusler compounds \cite{prediction_1,prediction_2}.

Despite a very strong experimental and theoretical interest to the
half-metallic ferromagnetism in Heusler alloys the  number  of the
studies of the  exchange  interactions  in Heusler alloys is still
very  small. The first contribution to the density  functional
theory of the exchange interactions in these systems was made in
an early paper by K\"ubler  \textit{et. al.}, \cite{Kubler2} where
the microscopic mechanisms of the magnetism of Heusler alloys were
discussed on the basis of the comparison of the ferromagnetic and
antiferromagnetic configurations of the  Mn moments. Recently, the
studies of the inter-atomic exchange interactions in several
Heusler compounds were reported by the present authors and
Kurtulus \textit{et al.} \cite{ersoy_1,ersoy_2,ersoy_3,Kurtulus}.

The  purpose of the present work is a first--principles study of
the exchange interactions and the temperature of magnetic phase
transition for four semi Heusler compounds NiCrZ (Z$=$P, Se, Te)
and NiVAs that were recently predicted to be half-metals
\cite{prediction_1,prediction_2}. For all four systems we find the
Curie temperature substantially exceeding room temperature. We
discuss the relation between the value of the Curie temperature
and half-metallicity. We demonstrate strong dependence of the
effective exchange interaction between 3\textit{d} atoms on the
\textit{sp} element (Z constituent).

The  paper is  organized as follows. In Sec. II  we present the
calculational approach. Section III contains the results of the
calculations and discussion. Section IV gives the
conclusions.

%---------------------------------------------------------------

\section{Computational Method}

The semi Heusler compounds crystallize in the $C1_b$--type structure.
The lattice consists of three interpenetrating fcc atomic sublattices.
Compared with full Heusler alloys the forth atomic sublattice is vacant.
In the calculations, the atomic positions of the fourth sublattice
are occupied by empty spheres.

The calculations are carried out with the augmented spherical
waves  method \cite{asw} within the atomic--sphere approximation
(ASA)\cite{asa}. The exchange--correlation potential is chosen in
the generalized gradient approximation \cite{gga}. A dense
Brillouin zone (BZ) sampling $30\times30\times30$ is used.

For each compound we performed calculations for two values of the
lattice parameter: the theoretical equilibrium parameter
\cite{prediction_1,prediction_2} and the lattice parameter of a
binary semiconductor (Table~\ref{tab:moments}) that can be
considered as a possible substrate for growing the corresponding
Heusler alloy: GaAs for NiCrP and NiCrSe and InP for NiVAs and
NiCrTe. The radii of all atomic spheres are chosen equal.

To calculate the interatomic exchange interactions we use the
frozen-magnon technique \cite{magnon} and map the results of
the calculation of the total energy of the helical magnetic
configurations
\begin{equation}
\label{spiral} {\bf s}_n=(\cos({\bf qR}_n)\sin{\theta}, \sin({\bf
qR}_n)\sin{\theta}, \cos {\theta})
\end{equation}
onto a classical Heisenberg Hamiltonian
\begin{equation}
\label{hamiltonian}
 H_{eff}=-  \sum_{i \ne j} J_{ij}
{\bf s}_i{\bf s}_j
\end{equation}
where $J_{ij}$ is an exchange interaction between two Cr (V) sites
and ${\bf s}_i$ is the unit vector pointing in the direction of
the magnetic moment at site $i$ , ${\bf R}_n$ are the lattice
vectors, ${\bf q}$ is the wave vector of the helix, $\theta$ polar
angle giving the deviation of the moments from the $z$ axis. Within
the Heisenberg model (\ref{hamiltonian}), the energy of frozen-magnon
configurations can be represented in the form
\begin{equation}
\label{eq:e_of_q} E(\theta,{\bf q})=E_0(\theta)+\sin^{2}\theta
J({\bf q})
\end{equation}
where $E_0$ does not depend on {\bf q} and $J({\bf q})$ is the the
Fourier transform of the  parameters of exchange  interaction
between pairs of  Cr(V) atoms:
\begin{equation}
\label{eq:J_q} J({\bf q})=\sum_{\bf R} J_{0{\bf R}}\:\exp(i{\bf
q\cdot R}).
\end{equation}

Calculating $ E(\theta,{\bf q})$ for a regular ${\bf q}$--mesh in
the Brillouin zone of the crystal and performing back Fourier
transformation one gets exchange parameters $J_{0{\bf R}}$ between
pairs of Cr(V) atoms.

The Curie temperature is estimated in the mean-field approximation
(MFA)
\begin{equation}
\label{eq:Tc_MFA}
k_BT_C^{MFA}=\frac{2}{3}\sum_{j\ne0}J_{0j}=\frac{M}{6\mu_B}\frac{1}{N}\sum_{\bf
q}\omega({\bf q})
\end{equation}
and random phase approximation (RPA)
\begin{equation}
\frac{1}{k_BT_C^{RPA}}=\frac{6\mu_B}{M}\frac{1}{N} \sum_{\bf q}
\frac{1}{\omega({\bf q})}
\end{equation}
where $ \omega({\bf q})=\frac{4}{M}[J(0)- J({\bf q})] $ is the
energy of  spin--wave excitations.

\begin{table}
\caption{Lattice parameters and magnetic moments (in
 $\mu_B$)  of NiVAs,  NiCrZ (Z$=$P, Se, Te) and NiMnSb. }
\begin{ruledtabular}
\begin{tabular}{llccccc}

$$&a(\AA)& Ni & V,Cr,Mn & Void  & Z & Cell
\\ \hline
NiVAs  &5.85                     & -0.02  & 2.05   & 0.06   & -0.09   & 2.0       \\
       &5.87$_{\textrm{(InP)}}$  & -0.03  & 2.06   & 0.06   & -0.09   & 2.0       \\

\hline
NiCrP  &5.59                      & -0.01  & 3.08   & 0.07  & -0.15   & 3.0       \\
       &5.65$_{\textrm{(GaAs)}}$  & -0.06  & 3.16   & 0.06  & -0.16   & 3.0       \\

\hline
NiMnSb$^{(a)}$ &5.64                      & 0.19  & 3.86   & 0.04   & -0.10   & 4.0      \\
\hline
NiCrSe &5.64                      & 0.24  & 3.64   & 0.12   & -0.01   & 4.0      \\
       &5.65$_{\textrm{(GaAs)}}$  & 0.23  & 3.65   & 0.12   & -0.01   & 4.0       \\

\hline
NiCrTe &5.84                      & 0.24   & 3.68  & 0.11   & -0.03   & 4.0       \\
       &5.87$_{\textrm{(InP)}}$   & 0.23   & 3.70  & 0.11   & -0.03   & 4.0       \\

\end{tabular}
\end{ruledtabular}
 \label{tab:moments}
\begin{flushleft}
$^{(a)}$  Ref.\onlinecite{ersoy_3} \\
\end{flushleft}
\end{table}

\section{Results and Discussion}

\begin{figure}[t]
\begin{center}
\includegraphics[scale=0.38]{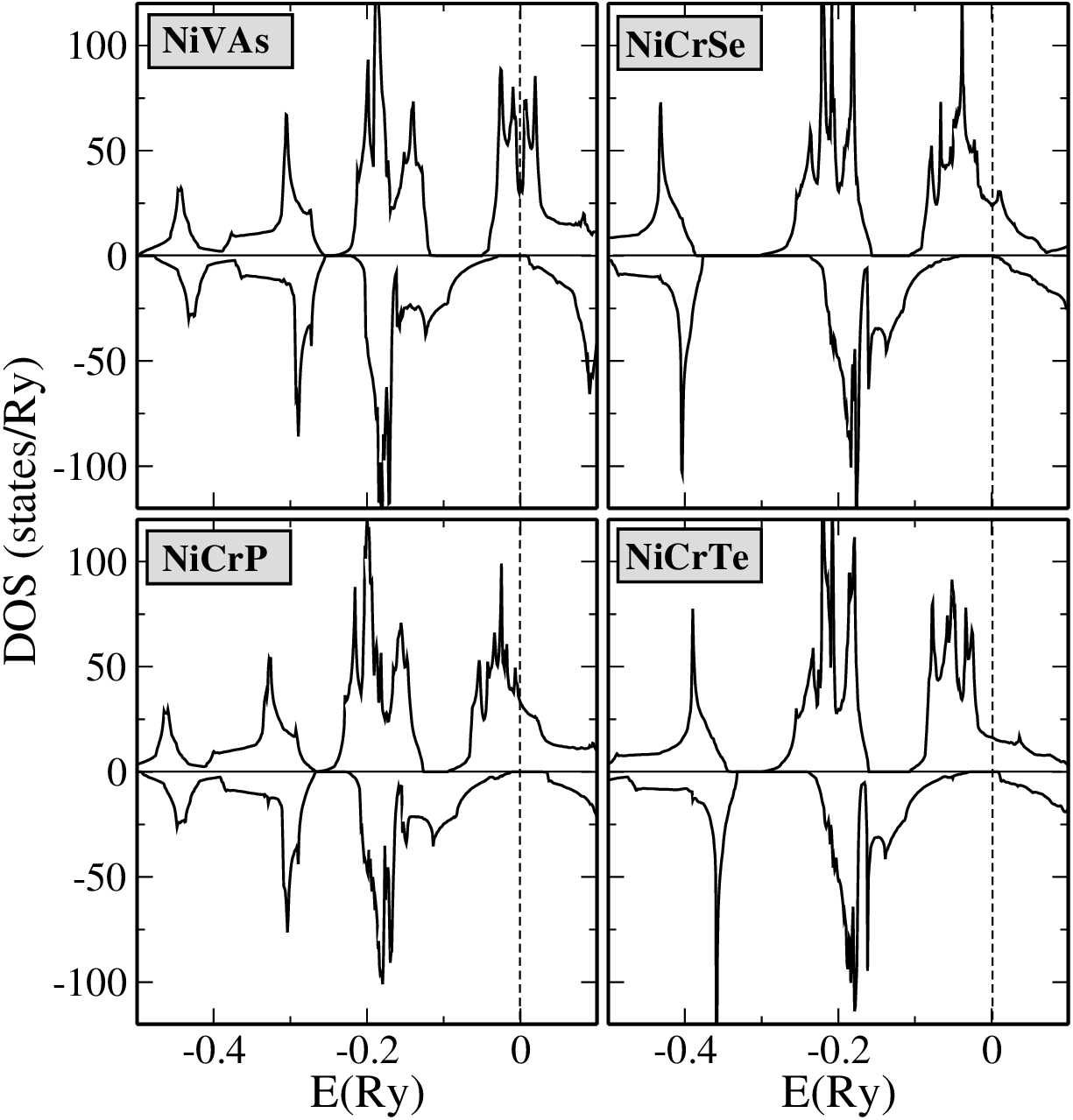}
\end{center}
\caption{Spin--projected total density of states of NiVAs and
NiCrZ  (Z=P, Se, Te)} \label{total_dos}
\end{figure}

\subsection{DOS and magnetic moments}

The two lattice parameters used in the calculations resulted for
all systems in very similar physical properties (see, e.g., Tables
\ref{tab:moments},\ref{tab:Tc}). Therefore the most of the results
will be presented for one lattice constant. All the discussion in
this section is valid for both lattice spacings.

In Fig.~\ref{total_dos}, we present the calculated electron
densities of states (DOS) of the ferromagnetic states of the four
Heusler compounds. All systems are found to be half-metallic with
the Fermi level lying in the semiconducting gap of the
minority-spin channel. Our DOS are in good agreement with  the DOS
presented in Refs.\cite{prediction_1,prediction_2}.

In Table \ref{tab:moments}, the calculated magnetic moments are
collected. Since the systems are half-metallic, the magnetic
moments per formula unit are integer: 2$\mu_B$ for NiVAs, 3$\mu_B$
for NiCrP and 4$\mu_B$ for NiCrSe and NiCrTe. The major part of
the magnetic moment comes from the second formula atom (V,Cr).
Small induced magnetic moments are found on Ni and $sp$ atoms.

\subsection{ Exchange constants and Curie temperature}

\begin{figure*}[t]
\begin{center}
\includegraphics[scale=0.46]{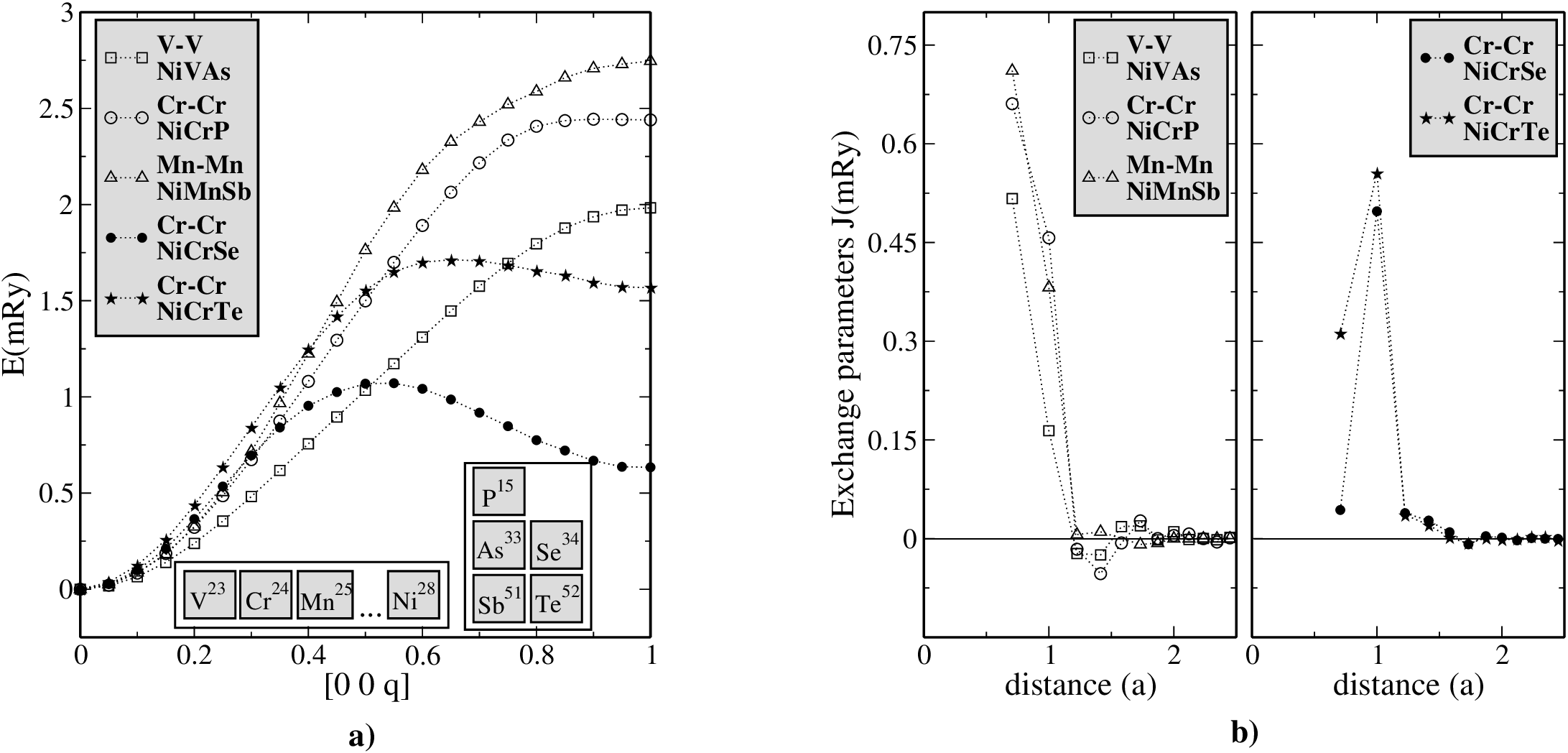}
\end{center}
\caption{a)Frozen--magnon  energies  as a function of the wave
vector ${\bf q}$ (in units of $2\pi/a$)  in NiVAs, NiMnSb and
NiCrZ (Z=P, Se, Te) for V--V, Mn--Mn and Cr--Cr interactions,
respectively. b)Interatomic  exchange interactions in NiCrZ
(Z$=$P, Se, Te) and NiVAs.  The distances are given in the units
of the lattice constant. The calculational data for NiVAs and
NiCrZ are presented for the theoretical equilibrium lattice
constant. The data for NiMnSb are from Ref. \cite{ersoy_3}. }
\label{exchange}
\end{figure*}

Figure~\ref{exchange}a presents the frozen-magnon dispersion for
one direction in the reciprocal space. Additionally to  NiCrZ
(Z$=$P, Se, Te) and NiVAs we present for comparison the results
for NiMnSb \cite{ersoy_3}. The five systems can be subdivided into
two groups. One group contains NiVAs, NiCrP and NiMnSb. Here the
frozen-magnon dispersions are monotonous and resemble, visually, a
simple cosinusoid. The second group contains NiCrSe and NiCrTe and
is characterized by nonmonotonous dispersions with a maximum close
to the center of the q-interval (Fig.~\ref{exchange}a). Note, that
the $sp$ elements (the third chemical-formula constituents) within
each of the group belong to the same column of the Mendeleev's
table whereas for different groups these columns are different.
The importance of the valency of the $sp$ element for magnetic
properties of Heusler alloys has been already observed in our
earlier studies \cite{ersoy_1}.

The calculated exchange parameters are given in
Fig.~\ref{exchange}b. Since the inter-atomic exchange parameters
are the Fourier transforms of the frozen-magnon dispersions they
reflect the properties of the dispersions: The exchange parameters
belonging to the same group show similar qualitative behavior. On
the other hand, there is strong difference between the systems
belonging to different groups (Fig.~\ref{exchange}b). In the first
group (Fig.~\ref{exchange}b, left panel), the strongest exchange
interaction takes place between nearest magnetic 3\textit{d}
atoms. This strongest interaction determines the cosinusoidal form
of the corresponding magnon dispersion. The sizable interaction
between the second-nearest magnetic 3\textit{d} atoms describes
the deviation of the dispersion from a simple cosinusoid.

In the second group of compounds (Fig.~\ref{exchange}b, right
panel), the strongest interaction is between the second-nearest
magnetic atoms. Because of the decreased period of the Fourier
component corresponding to the second exchange parameter the
dispersions are nonmonotonous and have the maximum not at the
boundary of the Brillouin zone but inside of it.

A remarkable   feature  of the   exchange interactions is their
short range character: the leading contribution into the Curie
temperature of all systems is provided by the interactions within
the first two coordination spheres. The interactions with further
coordination spheres are very weak and can be neglected in the
calculation of the Curie temperature. The interaction between Ni
atoms and  the interaction of Ni with V and Cr are very weak and
are not presented.

The calculated exchange parameters are used to evaluate the Curie
temperature (Table~\ref{tab:Tc}). It is important to note that the
similarity of the form of the magnon dispersions within one group
of compounds is not accompanied by a quantitative closeness of the
curves (Fig. \ref{exchange}a). Therefore the Curie temperature can
differ strongly for compounds belonging to the same group.

The Curie temperatures are estimated within two different schemes:
MFA and RPA. For all systems and for both theoretical schemes the
calculated Curie temperatures are substantially higher than room
temperature. The MFA always gives the value of $T_c$ that is larger than
the corresponding RPA value and usually overestimates the
experimental Curie temperature \cite{pajda,bouzerar}. The reason
for the difference of the MFA and RPA estimations is a different
weighting of the spin-wave excitations within two calculational
approaches: The MFA takes all excitations with the same weight
whereas the RPA gives a larger weight to the excitations with lower
energy. The RPA weighting is better grounded from the viewpoint of
statistical mechanics.

The analysis of the calculational data allows to make a number of
important conclusions. First, there is a strong influence of the
$sp$ atom on the value of $T_c$. Indeed, the comparison of NiCrSe
and NiCrTe that differ by the $sp$ atom shows that the Curie
temperature changes from about 500 K in NiCrSe to about 800 K in
NiCrTe. This strong dependence reveals the sensitivity of the
exchange interactions and the Curie temperature to the details of
the electron structure.

An interesting feature of the calculated Curie temperatures (Table
\ref{tab:Tc}) is a large difference between the MFA and RPA
estimations for the first group of compounds in contrast to a
small difference for the second group. Characterizing the relative
difference of the MFA and RPA values of the Curie temperature by
the relation $ [T_{c}^{MFA}-T_{c}^{RPA}]/T_{c}^{RPA}$ we get for
the first group of compounds a large value of 20-24\% compared to
a small value of 5-8\% in the second group. This feature reflects
the properties of the corresponding frozen-magnon spectra.

In MFA, the Curie temperature is determined by an arithmetic
average of the magnon energies while in RPA  $T_{c}$ is determined
by the harmonic average of the same quantities (the first average
is always larger than the second). In terms of magnon energies,
$T_{c}^{MFA}$ is equal to $T_{c}^{RPA}$ in the case that the
magnon spectrum is dispersion-less: the magnon energy does not
depend on the wave vector ${\bf q}$.

From Fig.~\ref{exchange} we see that the frozen--magnon curves of
the second group of compounds  are   flat in the second half of
the \textbf{q} interval demonstrating here a very weak dispersion.
On the other hand, the first group of compounds   have
considerable dispersion in this part of the \textbf{q} interval.
Comparing now the low-\textbf{q} parts of the curves we notice
that the curves of the second group of  compounds  lie higher than
the first group. Therefore, the relative contribution of the
low-energy magnons to the RPA value of the Curie temperature is
smaller in the second group of compounds.

This combination of features of the wave-vector dependencies of
the frozen-magnon energies is responsible for a larger (smaller)
difference between the RPA and MFA estimations of the Curie
temperature of the  first group (second group) of  compounds  in
the two cases.

%------------------------------------------------------------------------------
\subsection{Curie temperature and half-metallicity}

\begin{figure}[t]
\begin{center}
\includegraphics[scale=0.36]{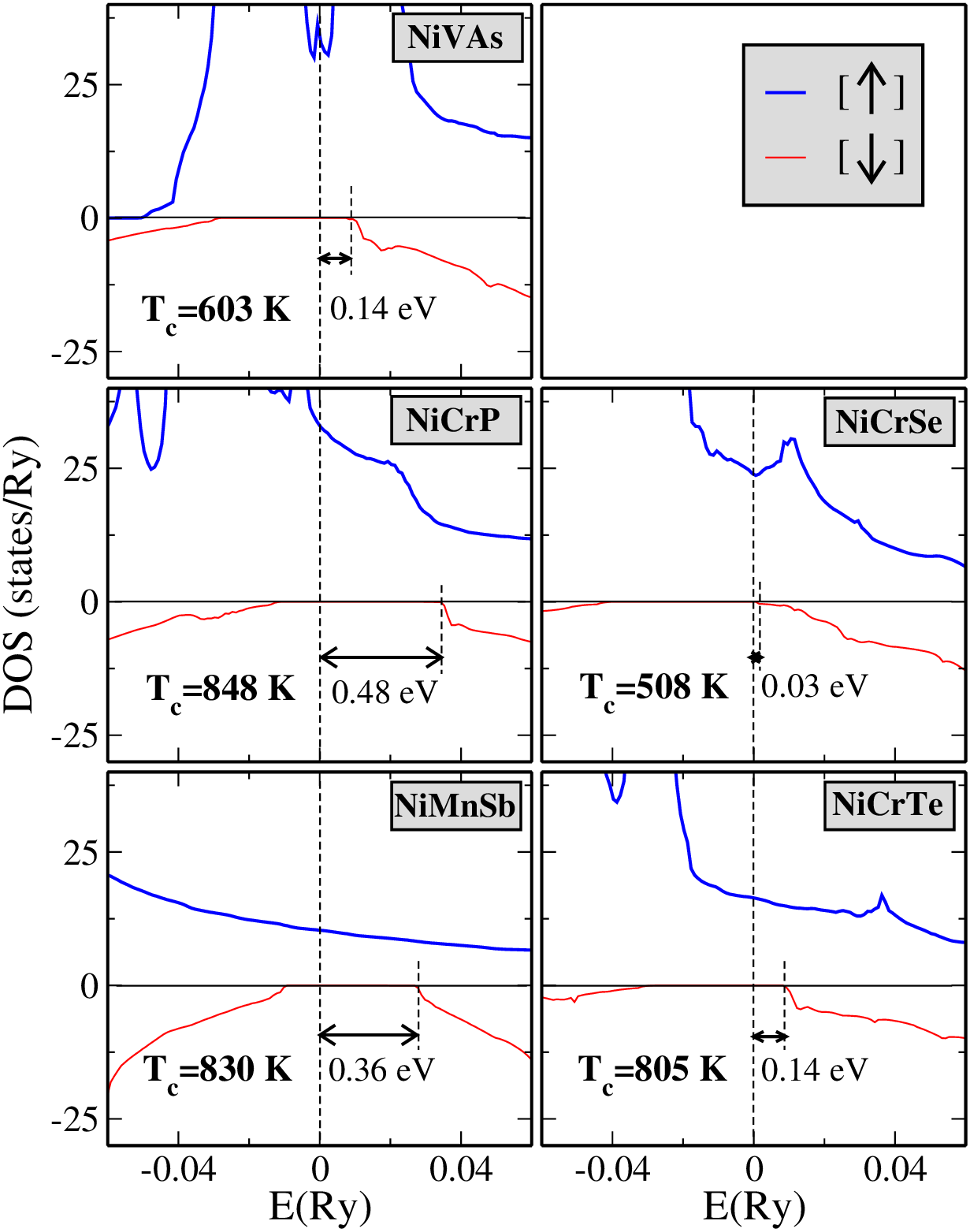}
\end{center}
\caption{Spin--projected total density of states of NiVAs, NiCrP,
NiMnSb, NiCrSe and NiCrTe  near Fermi level. Also shown are the
RPA estimations of the Curie temperature and  the energy distances
between the Fermi level and the bottom of the spin-down conduction
band.} \label{fermi_dos}
\end{figure}

An important question concerning the magnetism of the
half-metallic systems is the relation between half-metallicity and
Curie temperature. Indeed, a number of studies has shown that the
half-metallicity can stimulate an increase of the Curie
temperature \cite{kubler2002,Sakuma,sanyal}. The analysis of
Fig.~\ref{fermi_dos} allows us to establish a correlation between
the value of the Curie temperature and the energy distance
$\delta$ between the Fermi level and the upper edge of the
semiconducting gap. This quantity determines the spacing between
the highest occupied spin-up state and the lowest empty spin-down
state. For very small $\delta$ of 0.03 eV in NiCrSe we obtained
the lowest Curie temperature of 508 K. On the other hand, for
large  $\delta$ in NiCrP and NiMnSb we obtained the Curie
temperature substantially above 800 K.

Since the value of the Curie temperature is determined by  the
magnetic excitations to interpret the $\delta-T_c$ correlation we
need to understand the origin of the influence of the $\delta$
value on the spin-wave energies.

The magnon energies reflect the energy prize for the deviations of
the atomic moments from parallel directions
\cite{q,q2,superexchange}. In the ground-state ferromagnetic
configuration the spin-projection is a good quantum number and the
electron states with opposite spin projections do not interact.
The deviation of the atomic moments from parallel directions leads
to the mixing of the majority-spin and minority-spin states. The
hybridization of a pair of the states leads to their repulsion. As
a result, the energy of the lower state decreases (bonding state)
and the energy of the upper state increases (antibonding state).
If the lower state is occupied and the upper state is empty this
process leads to the decrease of the energy of the magnon making
ferromagnetic state energetically less favorable. The smaller is
the energy distance between interacting states the stronger is the
effect.

Coming back to the half-metallic compounds considered in the paper
we note that the hybridizational interaction of this type takes place
between the occupied majority-spin states below the Fermi level
and the empty spin-down states at the bottom of the conduction band.
The distance between these states is given by parameter $\delta$.
Therefore this process provides a mechanism for the correlation between
$\delta$ and Curie temperature.

Since the strength of the hybridizational repulsion increases with
decreasing energy distance between interacting states the negative
contribution to the spin-wave energies is larger in the case of
smaller $\delta$. However, the correlation between parameter
$\delta$ and $T_c$ or, more general, between the half-metallicity
and $T_c$ should not be considered as a universal rule. The
hybridizational repulsion considered above is only one of numerous
processes arising in a complex multi-band system with the
deviation of the atomic moments from the parallel directions. The
combined result of these processes cannot be predicted without the
direct calculation of the excitation energies. Such a calculation
must take into account the complexity of the electron structure of
a real system. Indeed, the comparison of NiVAs and NiCrTe shows
that both systems have the same $\delta$ of 0.14eV.  However,
their Curie temperatures differ strongly.

Also our numerical experiments with varying the lattice parameter
in half-metallic systems have shown that in some systems the loss
of the half-metallicity caused by this variation does not lead to
the decrease of the Curie temperature \cite{ersoy_3}.

\begin{table}
\caption{ MFA and RPA estimations of the Curie temperatures in
NiVAs, NiCrZ (Z$=$P, Se, Te) and NiMnSb. The third column
represent the relative difference of the  MFA and RPA estimations.
}
 \begin{ruledtabular}
\begin{tabular}{lccc}
$$   &    T$_c^{[MFA]}$(K) & T$_c^{[RPA]}$(K)&  T$_c$[$\frac{MFA-RPA}{RPA}$]
\\ \hline
NiVAs            &    723   &  603  & \% 20  \\
NiVAs$^{(a)}$    &    715   &  595  & \% 20 \\
\hline
NiCrP            &   1030   &  848  & \% 21 \\
NiCrP$^{(b)}$    &   938    &  770  & \% 22 \\
\hline
NiMnSb$^{(c)}$   &   1096    &  880  & \% 24 \\
\hline
NiCrSe          &   537    &  508  & \% 6  \\
NiCrSe$^{(b)}$  &   543    &  515  & \% 5  \\
\hline
NiCrTe          &   868    &  805  & \% 8 \\
NiCrTe$^{(a)}$  &   874    &  812  & \% 8 \\
\end{tabular}
\end{ruledtabular}
\begin{flushleft}
 \label{tab:Tc}
$^{(a)}$ lattice constant of GaAs \\
$^{(b)}$  lattice constant of InP \\
$^{(c)}$  Ref.\onlinecite{ersoy_3} \\
\end{flushleft}
\end{table}

\section{Conclusion}

We  have  systematically studied exchange interactions and Curie
temperature in predicted half--metallic semi Heusler compounds
NiVAs and NiCrZ (Z= P, Se, Te). The calculations are performed
within the parameter--free density functional theory.  The RPA and
MFA are used to estimate  the Curie temperatures. We show that the
behavior of exchange interactions in these systems vary strongly
depending on the Z constituent.

The exchange interactions are short range with the leading
contribution to the Curie temperature provided by the interactions
within the first two coordination spheres of the magnetic
3\textit{d} atoms. The predicted Curie temperatures of all four
systems are considerably higher than room temperature. The
relation between the half-metallicity and the value of the Curie
temperature is discussed. The combination of a high
spin-polarization of charge carriers and a high Curie temperature
makes these Heusler alloys interesting candidates for spintronics
applications. We hope that the present study will provide a
guideline for experimental work stimulating the fabrication of
these materials

\section*{Acknowledgements} The financial support of
Bundesministerium f\"ur Bildung und Forschung is acknowledged.

\end{document}